\nonstopmode

\newcommand{\Cal}[1]{{\cal #1}}
\newcommand{\ie}{i.e.{}}
\newcommand{\eg}{e.g.{}}
\newcommand{\X}[1]{_{\mathrm{#1}}}
\newcommand{\imag}{{\rm i}}
\newcommand{\euler}{\mathrm e}
\newcommand{\mat}[1]{\hbox{\boldmath{$#1$}\unboldmath}}
\newcommand{\Sum}{\sum\limits}
\newcommand{\Int}{\int\limits}
\newcommand{\differential}{\>\mathrm d}
\newcommand{\Lim}{\lim\limits}
\newcommand{\unitop}{\hat{\mathbbm{1}}}
\newcommand{\bra}[1]{\left<\right.\!#1\!\left.\right|}
\newcommand{\ket}[1]{\left|\right.\!#1\!\left.\right>}
\newcommand{\bracket}[2]{\left<\right.\!#1\left.\right|#2\!\left.\right>}
\newcommand{\nuclbracket}[3]{{}\X{E}\!\bra{#1} \otimes {}\X{N}\!\left<\right.
                             \!#2\left.\right|#3\!\left.\right>}
\newcommand{\XUV}{\textsc{xuv}}
\newcommand{\nbh}{\hbox{-}}
\newcommand{\xray}{x\nbh{}ray}
\newcommand{\Xray}{X\nbh{}ray}
\newcommand{\Tr}{\mathrm{Tr}\,}

\newcommand{\eref}[1]{(\ref{#1})}
\newcommand{\Eref}[1]{Equation~(\ref{#1})}
\newcommand{\sref}[1]{Sect.~\ref{#1}}

\documentclass[a4paper,10pt,twoside,5p,twocolumn]{elsarticle}
\journal{Chemical Physics}
\usepackage{bm,bbm,amsmath,amssymb,widetext}
\usepackage[nativepdf,bookmarks,bookmarksopen,bookmarksnumbered,raiselinks,%
breaklinks,%
pdftitle={Dissociating diatomic molecules in ultrafast and intense light},%
pdfauthor={Christian Buth},%
pdfsubject={Molecular Physics},%
pdfkeywords={density operator, master equation in Lindblad form,
quantum Monte Carlo algorithm, molecular dissociation,
ultrafast and intense light, quantum dynamics, Born-Oppenheimer
approximation, strong-field ionization, free electron laser,
attosecond light source, optical light, XUV light, x rays}]{hyperref}
\usepackage[anythingbreaks]{breakurl}
\allowdisplaybreaks

\DeclareMathOperator*{\SumInt}{%
\mathchoice%
  {\ooalign{$\displaystyle\sum$\cr\hidewidth$\displaystyle\int$\hidewidth\cr}}
  {\ooalign{\raisebox{.14\height}{\scalebox{.7}{$\textstyle\sum$}}\cr\hidewidth$\textstyle\int$\hidewidth\cr}}
  {\ooalign{\raisebox{.2\height}{\scalebox{.6}{$\scriptstyle\sum$}}\cr$\scriptstyle\int$\cr}}
  {\ooalign{\raisebox{.2\height}{\scalebox{.6}{$\scriptstyle\sum$}}\cr$\scriptstyle\int$\cr}}
}

\begin{document}
\begin{frontmatter}
\title{Dissociating diatomic molecules in ultrafast and intense light}
\author[uhd]{Christian Buth}
\ead{christian.buth@web.de}
\ead[url]{http://www.christianbuth.name}
\address[uhd]{Theoretische Chemie, Physikalisch-Chemisches Institut,
Ruprecht-Karls-Universit\"at Heidelberg,\linebreak
Im Neuenheimer Feld~229, 69120~Heidelberg, Germany}

\begin{abstract}
An \emph{ab initio} theory is devised for the quantum dynamics of
molecules undergoing (multiple) ionization in ultrafast and intense light.
Specifically, the intertwined problem of photoionization, radiative,
and electronic transitions in the course of dissociation is addressed
which arises, \eg, when molecules are exposed to \XUV~light or x~rays from free
electron lasers or attosecond light sources, but the approach
is equally useful in optical strong-field physics.
The coherent interaction of the molecule with the light in a specific
charge state is also treated.
I set out from an abstract formulation in terms of the quantum optical
notion of system-reservoir interaction using a master equation in
Lindblad form and analyze its short-time approximation.
First, I express it in a direct sum rigged Hilbert space for an efficient
solution with numerical methods for systems of differential equations.
Second, I derive a treatment via quantum Monte Carlo wave packet~(MCWP)
propagation.
The formalism is concretized to diatomic molecules in
Born-Oppenheimer approximation whereby molecular rotation is disregarded.
The numerical integration of the master equation is carried out with a
suitably factored density matrix that exploits the locality of the
Hamiltonian and the Lindblad superoperator with respect to the
internuclear distance.
The formulation of the MCWP for molecules requires a thorough
analysis of the quantum jump process;
namely, the dependence on the continuous distance renders a straight
wave packet promotion useless and, instead, a projected outer
product needs to be employed involving an integrated quantum jump operator.
\end{abstract}

\end{frontmatter}

\section{Introduction}

Understanding the interaction of molecules with ultrafast and intense light
poses a formidable challenge:
there is the coherent interaction of the molecular electronic states of a
specific charge state, photoionization, radiative, and electronic
transitions among charge states intertwined with the nuclear quantum
dynamics of rotation, vibration, and dissociation.
Seminal steps towards this issue were made experimentally by examining
the ion yields and the kinetic energy release~(KER) of nitrogen
molecules~(N$_2$) in x~rays of the Linac Coherent Light
Source~(LCLS) free electron laser~(FEL)~\cite{Hoener:FA-10,Fang:MI-12}.
Theoretically, phenomenological models~\cite{Buth:UA-12} and molecular rate
equations~\cite{Liu:RE-16} were devised to unravel the physics of
this multiple time-scale problem.
Recently, also a rate-equation description of the molecular
populations~\cite{Hao:EE-15,Inhester:WM-16} combined with
classical nuclear dynamics~\cite{Rudenko:FR-17} was used.

Nuclear dynamics in molecules in several charge states is treated in a
time-dependent view on quantum mechanics~\cite{Tannor:IQ-07}.
Foundational work on the interdependence of electronic decay
and nuclear dynamics has been carried out in Refs.~\cite{Cederbaum:ND-93,%
Cederbaum:SD-93,Pahl:RD-99,Pahl:NB-99,Scheit:IP-03,Cederbaum:RA-11,%
Sisourat:ND-13}.
Even optical laser control of the dissociation of~N$_2^{2+}$
can be realized~\cite{Glownia:TR-10,Hanna:LC-17}.
When emitted photons and electrons are not considered, an open
quantum-system approach based on a master equation
in Lindblad form~\cite{Lindblad:QG-76,Louisell:QS-90,Blum:DM-96,Scully:QO-97,%
Meystre:QO-99,Gardiner:QN-04} needs to be employed.
Such a description has been developed for diatomic molecules in
Refs.~\cite{Leth:MI-09,Leth:MC-09,Leth:MC-10,Leth:DD-10,%
Leth:DM-11,Leth:DI-11,Jing:MC-16,Jing:LI-16}
using Monte Carlo wave packet propagation~(MCWP), a quantum Monte Carlo~(QMC)
formalism~\cite{Dalibard:DP-92,Molmer:MC-93,Molmer:MC-96,Plenio:QJ-98,%
Meystre:QO-99,Gardiner:QN-04}.
Related notions of a treatment of dissipative processes have
been named quantum-jump and quantum-trajectory approach~\cite{Plenio:QJ-98}.
The KER of H$_2$, HD, and D$_2$~molecules
in strong optical fields~\cite{Leth:MI-09,Leth:MC-09,Leth:MC-10,Leth:DD-10,%
Leth:DI-11,Jing:MC-16,Jing:LI-16}
and of~O$_2$ in \XUV~light of the SPring-8 Compact SASE
Source~(SCSS) FEL~\cite{Leth:DM-11,Leth:DI-11,Fukuzawa:IM-09}
were investigated and a description via incoherent ionization rates
was shown to be sufficient.
Likewise, Refs.~\cite{Buth:UA-12,Liu:RE-16} indicate that also
radiative and electronic decay can be described by incoherent rates.
This reduces the intricacy of the problem because now only
the coherent interaction with light and the nuclear quantum dynamics
need to be fully accounted for.
Such coherent effects manifest in photoexcitation with the ensuing quantum
dynamics which has been studied for a long time, \eg,
Ref.~\cite{Beck:MC-00},
and is described by the present formalism as well as a special case.
Yet there are issues with the definition of the quantum jump
operator in Refs.~\cite{Leth:MI-09,Leth:MC-09,Leth:MC-10,Leth:DD-10,%
Leth:DM-11,Leth:DI-11,Jing:MC-16,Jing:LI-16} which lead to an incorrect
MCWP scheme there despite a convincing agreement between computations and
experiment.

Solving the master equation is in several ways simplified by MCWP.
First, a density matrix has a quadratic dependence on the number
of basis states of the rigged Hilbert space~\cite{Madrid:RH-05,Bohm:BS-09}
employed while a wave packet has only a linear dependence.
Second, the propagation of the density matrix is inseparable and
needs to be carried out as a whole whereas the propagation of wave
packets in MCWP can be performed in parallel.
Specifically, if the system is initially in a probabilistic mixture,
MCWP propagation can be done for sufficiently many starting wave packets to
adequately sample the mixture~\cite{Meystre:QO-99}.
Third, MCWP offers an insightful physical interpretation of the
interaction of the system with light in contrast to the abstract
density operator formalism.
Namely, if the ejection of photons and electrons from the molecule can be
measured by gedanken detectors, then MCWP describes the physical
quantum jumping between states and not only represents
a sophisticated theoretical device to solve the master
equation~\cite{Dalibard:DP-92,Molmer:MC-93,Molmer:MC-96,Plenio:QJ-98,%
Meystre:QO-99,Leth:MI-09,Leth:MC-09,Leth:MC-10,Leth:DD-10,%
Leth:DM-11,Leth:DI-11,Jing:MC-16,Jing:LI-16}.

The article is organized as follows.
The master equation in Lindblad form is discussed abstractly
in~\sref{sec:ctqmc}, first, in general [\sref{sec:reservoir}] and,
second, in a short-time approximation [\sref{sec:shorttime}].
Assuming a system in various charge states in~\sref{sec:directsum},
the master equation is expressed in a direct sum~\cite{Tsalenko:DS-14}
rigged Hilbert space~\cite{Madrid:RH-05,Bohm:BS-09}.
A discrete time quantum Monte Carlo algorithm is devised in~\sref{sec:MCWP}.
The formalism of sections~\ref{sec:ctqmc} is concretized to dissociating
molecules in~\sref{sec:theory}.
The direct sum rigged Hilbert space, the Hamiltonian, and the master equation
are explicated in~\sref{sec:molhilbert}, \sref{sec:hamiltonian}, and
\sref{sec:molmaster}, respectively.
The density matrix is factored in~\sref{sec:factordens} with respect to the
internuclear distance and a system of partial differential equations~(PDEs)
is derived which is amenable to a solution with established numerical methods.
The MCWP scheme is devised in~\sref{sec:molQMC} for an integrated
quantum jump operator.
Conclusions are drawn in~\sref{sec:conclusion}.
I formulate the theory for diatomic molecules which, however, can be
generalized to larger molecules, \eg, by using multiconfiguration time-dependent
Hartree~(MCTDH)~\cite{Meyer:MC-90,Manthe:WD-92,Beck:MC-00} which is
a powerful approach to treat nuclear dynamics quantum mechanically for a
specific charge state of a molecule.
Atomic units are used throughout~\cite{Szabo:MQC-89,Hartree:WM-28}.

\section{Master equation in Lindblad form}
\label{sec:ctqmc}
\subsection{System-reservoir interaction}
\label{sec:reservoir}

The most general quantum mechanical description of a system is
given by a density operator or probabilistic operator%
\footnote{The density operator is frequently also referred
to as ``statistical operator'' which is a misnomer as the density
operator models a random phenomenon and it is not used
for the analysis of data from such a phenomenon.}
which is denoted by~$\hat \rho(t)$ for time~$t$~\cite{Louisell:QS-90,%
Blum:DM-96,Scully:QO-97,Merzbacher:QM-98,Meystre:QO-99,Gardiner:QN-04,%
Cohen-Tannoudji:QM-05}.
It facilitates to jointly describe pure quantum states and probabilistic
mixtures where for the latter, insufficient information is available,
\ie, the state of the system has not been fully characterized in terms
of a complete experiment, and thus it cannot be expressed in terms of
a single state vector.
The behavior of~$\hat \rho(t)$ is elucidated within the quantum optical
notion of system-reservoir interaction~\cite{Louisell:QS-90,Blum:DM-96,%
Scully:QO-97,Meystre:QO-99,Gardiner:QN-04} as theoretical framework.
It is particularly useful, in my case of an open system, for which
electrons and photons are ejected into the continuum, \ie, the
reservoir, and thus particles and energy are dissipated.
As the state of the reservoir remains undetected (there are only gedanken
detectors), a single-state-vector representation is not feasible.

The time-dependent quantum dynamics of the system and the reservoir is
described by the analog of the Schr\"odinger equation for the joint density
operator which is termed the Liouville or von Neumann equation;
it describes physical processes which are reversible in
time~\cite{Louisell:QS-90,Blum:DM-96,Scully:QO-97,%
Meystre:QO-99,Gardiner:QN-04}.
The combined system and reservoir is a closed system;
the master equation for the system only is derived by considering the
system-reservoir coupling up to second order in perturbation theory.
Allowing the reservoir to ``loose its memory'' by introducing the
Markoff approximation---which is in this context equivalent to
Weisskopf-Wigner theory~\cite{Weisskopf:LB-30} of decay
processes~\cite{Meystre:QO-99}---results in irreversibility of a quantum
process.
This leads to an equation for the reduced density operator that
comprises only the system reading
\begin{equation}
  \label{eq:mastereqabs}
  \dfrac{\partial \hat \rho(t)}{\partial t} = -\imag \, [\hat H, \hat \rho(t)]
    + \hat{\Cal L}\X{Ld}[\hat \rho(t)] \; ,
\end{equation}
with the Hamiltonian~$\hat H$ of the system.
The bracket in the first term on the right-hand side of the equation denotes a
commutator which accounts for the coherent evolution of the system;
without system-reservoir interaction, this is the only term on the right-hand
side and Eq.~\eref{eq:mastereqabs} has the form of a Liouville equation.
The second term is the Liouvillian
superoperator~$\hat{\Cal L}\X{Ld}[\hat \rho(t)]$ that describes the
dissipation of electrons and photons into the continuum.
Additional norm nonconserving terms arise, if not all final states of the
system are incorporated because
then transitions to states not included in the density operator
cause a loss of norm.
I assume that all final states are incorporated and thus the trace of the
density operator is unity for all times.
This property of~$\hat{\Cal L}\X{Ld}[\hat \rho(t)]$ implies that the
Liouvillian has Lindblad form~\cite{Lindblad:QG-76,Louisell:QS-90,%
Blum:DM-96,Scully:QO-97,Meystre:QO-99,Gardiner:QN-04}.
It is expressed in terms of quantum jump operators
\begin{equation}
  \label{eq:jumper}
  \hat J_{\phi \psi} = \sqrt{\gamma_{\phi \psi}(t)} \> \ket{\phi} \bra{\psi}
    \; ,
\end{equation}
for~$\phi, \psi \in \mathbb B$ with a basis~$\mathbb B$ of the rigged
Hilbert space~\cite{Madrid:RH-05,Bohm:BS-09}.
The transition rate from~$\ket{\psi}$ to~$\ket{\phi}$
is~$\gamma_{\phi \psi}(t)$;
it vanishes for~$\phi = \psi$.
Thus norm conserving relaxation processes~\cite{Lindblad:QG-76,%
Louisell:QS-90,Blum:DM-96,Scully:QO-97,Meystre:QO-99,Gardiner:QN-04}
are effected by
\begin{eqnarray}
  \label{eq:Lindblad}
  \hat{\Cal L}\X{Ld}[\hat \rho(t)] &=& \hat{\Cal L}\X{Nm}[\hat \rho(t)]
    + \hat{\Cal L}\X{Jp}[\hat \rho(t)] \\
  &=& - \dfrac{1}{2} \SumInt_{\phi, \psi \in \mathbb B}
    [ \hat J_{\phi \psi}^{\dagger} \hat J_{\phi \psi}^{\phantom{\dagger}} \,
    \hat \rho(t) + \hat \rho(t) \, \hat J_{\phi \psi}^{\dagger}
    \hat J_{\phi \psi}^{\phantom{\dagger}} ] \nonumber \\
  &&{} + \SumInt_{\phi, \psi \in \mathbb B} \hat J_{\phi \psi}
    ^{\phantom{\dagger}} \, \hat \rho(t) \, \hat J_{\phi \psi}^{\dagger}
    \nonumber \\
  \label{eq:Lindbladabs}
  &=& - \dfrac{1}{2} \SumInt_{\psi \in \mathbb B}
    \Gamma_{\psi}(t) \> [ \hat{\Cal P}_{\psi} \, \hat \rho(t) +
    \hat \rho(t) \, \hat{\Cal P}_{\psi} ] \\
  &&{} + \SumInt_{\phi, \psi \in \mathbb B} \gamma_{\phi \psi}(t) \bra{\psi}
    \hat \rho(t) \ket{\psi} \, \hat{\Cal P}_{\phi} \nonumber \; ;
\end{eqnarray}
it consists of a norm reducing superoperator~$\hat{\Cal L}\X{Nm}[\hat
\rho(t)]$ and a quantum jump superoperator~$\hat{\Cal L}\X{Jp}[\hat \rho(t)]$.
The total transition rate is
\begin{equation}
  \label{eq:MCtotdecra}
  \Gamma_{\psi}(t) = \SumInt_{\phi \in \mathbb B} \gamma_{\phi \psi}(t) \; .
\end{equation}
The projector on a state vector~$\ket{\psi}$ is~$\hat{\Cal P}_{\psi}
= \ket{\psi} \bra{\psi}$.
In the brackets in Eq.~\eref{eq:Lindbladabs} an anticommutator
of~$\hat \rho(t)$ with~$\hat{\Cal P}_{\psi}$ is spelled out.

The diagonal elements of the density matrix~$p_{\phi}(t) =
\bra{\phi} \hat \rho(t) \ket{\phi}$ are the populations---\ie,
probabilities for discrete states and probability densities
for continuum states---to find the system in the
state~$\ket{\phi}$ for~$\phi \in \mathbb B$.
Taking the diagonal elements of the master
equation~\eref{eq:mastereqabs}, \eref{eq:Lindblad}, I obtain a Pauli master
equation~\cite{Pauli:FS-28,Louisell:QS-90,Blum:DM-96}---a rate
equation---reading
\begin{eqnarray}
  \label{eq:PauliMaster}
  \dfrac{\partial p_{\phi}(t)}{\partial t} &=& -\imag \,
    \bra{\phi} \bigl[ \hat H, \hat \rho(t) \bigr]
    \ket{\phi} + \SumInt_{\psi \in \mathbb B} \gamma_{\phi \psi}(t) \>
    p_{\psi}(t) \nonumber \\
  &&{} - \Gamma_{\phi}(t) \> p_{\phi}(t) \; .
\end{eqnarray}
The first term on the right-hand side describes the coherent time
evolution of the system governed by~$\hat H$ whereas the remaining
summands are relaxation terms from system-reservoir
interaction~\eref{eq:Lindblad}.
Rate equations are frequently applied in the semiclassical
theory of the laser~\cite{Scully:QO-97,Meystre:QO-99}
to describe the absorption and stimulated emission of light.
If~$\hat H$ has a matrix representation that is diagonal
with respect to~$\mathbb B$, \ie, it does not couple basis states,
the matrix element of the commutator in~\eref{eq:PauliMaster} vanishes.
In this form, rate equations play an important role in the understanding
of the ionization of atoms~\cite{Buth:UA-12,Buth:NX-18,Obaid:FL-18,Buth:LT-19}
and molecules~\cite{Buth:UA-12,Liu:RE-16} by intense
light in the optical and \xray~regimes.
The derivation of Eq.~\eref{eq:PauliMaster} from Eqs.~\eref{eq:mastereqabs},
\eref{eq:Lindblad} represents an \emph{a posteriori} justification
of the rate-equation approximation of
Refs.~\cite{Buth:UA-12,Liu:RE-16,Buth:NX-18,Obaid:FL-18,Buth:LT-19}.
This, in turn, assures me that the present theory describes the interaction
with the light adequately.

\subsection{Short-time approximation}
\label{sec:shorttime}

I recast the master equation~\eref{eq:mastereqabs} with the Lindblad
superoperator~\eref{eq:Lindblad} into the form~\cite{Meystre:QO-99} of
\begin{equation}
  \label{eq:mastereqMCWP}
  \dfrac{\partial \hat \rho(t)}{\partial t}
    = -\imag \, \bigl( \hat H^{\phantom{\dagger}}\X{eff} \, \hat \rho(t)
    - \hat \rho(t) \, \hat H^{\dagger}\X{eff} \bigr)
    + \hat{\Cal L}\X{Jp}[\hat \rho(t)] \; ,
\end{equation}
with the effective Hamiltonian
\begin{equation}
  \label{eq:HamMC}
  \hat H\X{eff} = \hat H + \hat V\X{ra} \, ,
\end{equation}
that comprises the Hermitian~\cite{Merzbacher:QM-98}
Hamiltonian~$\hat H = \hat H^{\dagger}$
for the system and the non-Hermitian decay operator
for the transitions away from a state~$\ket{\psi}$ that cause loss
of population of this state with the rate~$\Gamma_{\psi}(t)$ via
\begin{equation}
  \label{eq:Vraabs}
  \hat V\X{ra} = - \dfrac{\imag}{2} \SumInt_{\psi \in \mathbb B}
    \Gamma_{\psi}(t) \> \hat{\Cal P}_{\psi} \; .
\end{equation}
Inserting the expression for~$\hat H\X{eff}$ from Eq.~\eref{eq:HamMC},
\eref{eq:Vraabs} into Eq.~\eref{eq:mastereqMCWP}, the
sum of the superoperator for the coherent evolution~$-\imag \, [\hat H,
\hat \rho(t)]$ and the norm-reducing superoperator~$\hat{\Cal L}\X{Nm}
[\hat \rho(t)]$ are obtained such that only~$\hat{\Cal L}\X{Jp}
[\hat \rho(t)]$ remains to be added to reproduce the master
equation~\eref{eq:mastereqabs}, \eref{eq:Lindblad}.

The master equation~\eref{eq:mastereqMCWP}
can be approximately integrated for a short time interval~$\delta t$ by
replacing the temporal derivative on the left-hand side of
Eq.~\eref{eq:mastereqMCWP} by a difference quotient yielding
\begin{eqnarray}
  \label{eq:mastereqDeltat}
  \hat \rho(t + \delta t) &=& \hat \rho(t) - \imag \, \delta t \,
    \bigl( \hat H\X{eff} \, \hat \rho(t) - \hat \rho(t) \,
    \hat H^{\dagger}\X{eff} \bigr) \\
  &&{} + \delta t \, \hat{\Cal L}\X{Jp}[\hat \rho(t)] + O(\delta t^2) \; .
    \nonumber
\end{eqnarray}
The Landau symbol~\cite{Shabunin:OR-11} big-$O$ indicates terms
with~$\delta t^2$ and higher orders.
I refer to this equation as short-time master equation.
The separation into contributions from~$\hat H\X{eff}$ and
contributions from~$\hat{\Cal L}\X{Jp}[\hat \rho(t)]$ is
fundamental and a consequence of the linearization of the
master equation with respect to time differences~$\delta t$
by neglecting the terms~$O(\delta t^2)$.
The time interval~$\delta t$ needs to be longer than the reservoir
memory time such that the Markoff approximation can be
made;
but $\delta t$ must also be shorter than the system time scale such that
there is \emph{no} appreciable change in the system
variables~\cite{Meystre:QO-99}.

I find from Eq.~\eref{eq:mastereqDeltat} by taking the
trace on both sides of the equation and using the linearity of the
trace~\cite{Louisell:QS-90,Blum:DM-96,Merzbacher:QM-98,Cohen-Tannoudji:QM-05}
and the fact that the trace of the probabilistic operator is unity at all
times, that the trace of the two terms before~$O(\delta t^2)$ are the
same in magnitude with opposite sign such that they cancel upon neglecting
terms of~$O(\delta t^2)$, \ie,
\begin{eqnarray}
  \label{eq:tracecond}
  \delta p(t) &=& \imag \, \delta t \, \Tr \bigl(
    \hat H^{\phantom{\dagger}}\X{eff} \, \hat \rho(t)
    - \hat \rho(t) \, \hat H^{\dagger}\X{eff} \bigr) \\
  &=& \delta t \; \Tr (\hat{\Cal L}\X{Jp}[\hat \rho(t)])
    + O(\delta t^2) \; . \nonumber
\end{eqnarray}
The Eq.~\eref{eq:tracecond} describes the fact that the non-Hermitian terms in
Eq.~\eref{eq:HamMC} reduce the norm of the wave packet with time, \ie, the
probability to remain in a state decreases.
The loss of norm in the course of the temporal evolution
is only apparent because of quantum jumps to other states which
counter the loss and cause the conservation of the trace of the
probabilistic operator.

The density operator~$\hat \rho(\tau)$ at the initial time~$\tau$
describes a probabilistic mixture of states and is represented in
terms of orthonormal state vectors~$\ket{\Psi, \tau}$
for~$\Psi \in \mathbb S$.
The countable set~$\mathbb S \subseteq \mathbb B$ is the initial-state
composition of the density operator;
for an initial state which is pure, $\mathbb S$~has
only one element, \ie, the system is described by a single state vector.
The probabilities to find the system in the respective state are~$W_{\Psi}$.
As remains to be shown, $\hat \rho(t)$~is diagonal at all times~$t$
in terms of the time-evolving state vectors of~$\mathbb S$.
Let the density operator be diagonal at
time~$t$~\cite{Blum:DM-96,Meystre:QO-99}, \ie,
\begin{equation}
  \label{eq:gendensexp}
  \hat \rho(t) = \Sum_{\Psi \in \mathbb S} \ket{\Psi, t} W_{\Psi}
    \bra{\Psi, t} \; .
\end{equation}
Specifically, the trace condition~\eref{eq:tracecond} yields---for the
probability of the system to undergo a quantum jump in the time
interval~$\delta t$,---by inserting Eq.~\eref{eq:gendensexp}, the expression
\begin{eqnarray}
  \label{eq:lossnorm}
  \delta p(t) &=& \Sum_{\Psi \in \mathbb S} W_{\Psi} \> \delta p_{\Psi}(t)
    \nonumber \\
  &=& \imag \, \delta t \, \Sum_{\Psi \in \mathbb S} W_{\Psi} \bra{\Psi, t}
    \hat H^{\phantom{\dagger}}\X{eff} - \hat H^{\dagger}\X{eff} \ket{\Psi, t} \\
  &=& \delta t \Sum_{\Psi \in \mathbb S} W_{\Psi} \SumInt_{\psi \in
    \mathbb B} \Gamma_{\psi}(t) \> |\bracket{\psi}{\Psi, t}|^2 \; , \nonumber
\end{eqnarray}
with the total decay rate~\eref{eq:MCtotdecra} of~$\ket{\psi}$.

Inserting Eq.~\eref{eq:gendensexp} into Eq.~\eref{eq:mastereqDeltat},
I can express the first and second terms on the right-hand side
of Eq.~\eref{eq:mastereqDeltat}, \ie, without the quantum jump
contribution, in terms of the time evolution of a state vector%
\footnote{In stating Eq.~\eref{eq:schromcDeltat}, I do not presuppose that the
temporal evolution is given by~$\ket{\tilde \Psi, t + \delta t}
= \euler^{-\imag \, \hat H\X{eff} \, \delta t} \ket{\Psi, t}$
which is only fulfilled for a time-independent~$\hat H\X{eff}$.}
obeying
\begin{equation}
  \label{eq:schromcDeltat}
  \ket{\tilde \Psi, t + \delta t} = \ket{\Psi, t} - \imag \, \delta t \,
    \hat H\X{eff} \ket{\Psi, t} + O(\delta t^2) \; .
\end{equation}
This is a short-time approximation of the temporal propagation
of a wave packet~\cite{Dalibard:DP-92,Molmer:MC-93,Molmer:MC-96,%
Plenio:QJ-98,Meystre:QO-99} using the time-dependent Schr\"odinger
equation~\cite{Merzbacher:QM-98} with the effective
Hamiltonian~\eref{eq:HamMC}, \ie,
\begin{equation}
  \label{eq:schromc}
  \imag \, \dfrac{\partial}{\partial t} \ket{\Psi, t} = \hat H\X{eff}
    \ket{\Psi, t} \; .
\end{equation}
I use a tilde on~$\ket{\tilde \Psi, t + \delta t}$ in
Eq.~\eref{eq:schromcDeltat} to indicate that---due to the non-Hermiticity
of~$\hat H\X{eff}$---it is not normalized but
\begin{eqnarray}
  \label{eq:normNHprop}
  && \bracket{\tilde \Psi, t + \delta t}{\tilde \Psi, t + \delta t} \nonumber \\
  &=& 1 - \imag \, \delta t \bra{\Psi, t} \hat H^{\phantom{\dagger}}\X{eff}
    - \hat H^{\dagger}\X{eff} \ket{\Psi, t} + O(\delta t^2) \\
  &=& 1 - \delta p_{\Psi}(t) + O(\delta t^2) \; , \nonumber
\end{eqnarray}
holds with~\eref{eq:lossnorm}.
Likewise inserting Eq.~\eref{eq:gendensexp} into Eq.~\eref{eq:mastereqDeltat},
I express the third term on the right-hand side as a double sum over the
outer product of~$\sqrt{\delta t} \, \hat J_{\phi\psi} \ket{\Psi, t}$.
This state is also not normalized but I have
\begin{equation}
  \label{eq:normNHjump}
  \delta p_{\Psi}(t) = \delta t \SumInt_{\phi, \psi \in \mathbb B}
    \bra{\Psi, t} \hat J_{\phi\psi}^{\dagger} \hat J_{\phi\psi}
    ^{\phantom{\dagger}} \ket{\Psi, t} \; ,
\end{equation}
with Eqs.~\eref{eq:jumper} and \eref{eq:lossnorm}.

I rewrite the short-time master equation~\eref{eq:mastereqDeltat} based on
probabilities~\cite{Meystre:QO-99} using Eqs.~\eref{eq:gendensexp},
\eref{eq:schromcDeltat}, \eref{eq:normNHprop}, and \eref{eq:normNHjump}
arriving at
\begin{widetext}
\begin{eqnarray}
  \label{eq:mastereqProb}
  \hat \rho(t + \delta t) &=& \Sum_{\Psi \in \mathbb S} W_{\Psi} \>
    (1 - \delta p_{\Psi}(t)) \, \dfrac{\ket{\tilde \Psi, t + \delta t}}
    {\sqrt{1 - \delta p_{\Psi}(t)}} \dfrac{\bra{\tilde \Psi, t + \delta t}}
    {\sqrt{1 - \delta p_{\Psi}(t)}} \\
  &&{} + \Sum_{\Psi \in \mathbb S} W_{\Psi} \> \delta p_{\Psi}(t) \,
     \SumInt_{\phi, \psi \in \mathbb B} \sqrt{\dfrac{\delta t}{\delta
     p_{\Psi}(t)}} \hat J_{\phi\psi}^{\phantom{\dagger}} \ket{\Psi, t}
     \bra{\Psi, t} \hat J_{\phi\psi}^{\dagger} \sqrt{\dfrac{\delta t}
     {\delta p_{\Psi}(t)}} + O(\delta t^2) \; . \nonumber
\end{eqnarray}
\end{widetext}
The probabilistic operator at time~$t + \delta t$ is diagonal with respect
to the new states;
it is composed of a non-Hermitian propagation from the first and
second terms on the right-hand side of Eq.~\eref{eq:mastereqDeltat}
and a quantum-jump contribution from the third term in
Eq.~\eref{eq:mastereqDeltat} which
are multiplied by the remaining norm~$1 - \delta p_{\Psi}(t)$
[Eq.~\eref{eq:normNHprop}] and the loss of norm~$\delta p_{\Psi}(t)$
[Eq.~\eref{eq:normNHjump}], respectively.

\subsection{Numerical integration}
\label{sec:directsum}

The master equation~\eref{eq:mastereqMCWP} is a system of coupled
differential equations which are first-order in the time derivative.
The short-time master equation~\eref{eq:mastereqDeltat} has the form of
a numerical scheme for a discretized time upon neglecting terms~$O(\delta t^2)$
and can be used right away.
This is actually the Euler method~\cite{Vapnyarskii:EM-11} in the case
that the Hamiltonian does not contain any differential operators and
thus~\eref{eq:mastereqMCWP} is a system of first-order ordinary
differential equations.
In practice, however, one would not employ the Euler
method~\cite{Vapnyarskii:EM-11} but rather use, \eg,
the more accurate Runge-Kutta method of fourth order~\cite{Bobkov:RK-12}
as done in Ref.~\cite{Buth:LA-08}
to integrate the Liouville equation for the density operator of
molecules aligned by an intense optical laser.

Although a method for solving~\eref{eq:mastereqMCWP} is introduced
in the previous paragraph, further simplifications can be achieved
by making more assumptions about the context which is here photoionization,
radiative, and electronic transitions whereby the photons and electrons
liberated in transitions induced by the Lindblad superoperator are disregarded.
Let the neutral system have~$N \in \mathbb N$~electrons and
let the maximum charge state considered be~$N\X{el} \in \mathbb N_0$,
$N\X{el} \leq N$.
Then the states in~$\mathbb B$ are a basis of the direct
sum~\cite{Tsalenko:DS-14} of the $n$-electron rigged Hilbert
spaces~\cite{Madrid:RH-05,Bohm:BS-09} for~$n \in \{ N - N\X{el}, \ldots, N \}
= \mathbb A\X{el}$.
The equations of the previous~\sref{sec:ctqmc} are formulated in this space.
The Hamilton operator decomposes~$\hat H = \Sum_{n \in \mathbb A\X{el}}
\hat H^{(n)}$ into a sum of Hamiltonians~$\hat H^{(n)}$ for charge state~$n$.
Looking at~$\hat V\X{ra}$ [Eq.~\eref{eq:Vraabs}], I realize, that the
sum/integral therein can be partitioned such that
$\hat H\X{eff} = \Sum_{n \in \mathbb A\X{el}}
\hat H^{(n)}\X{eff}$ holds with~\eref{eq:HamMC}.
Let~$\hat{\mathfrak P}^{(n)}$ be the projection operator onto the $n$-electron
rigged Hilbert space and $\unitop = \Sum_{n \in \mathbb A\X{el}}
\hat{\mathfrak P}^{(n)}$.
Looking at Eq.~\eref{eq:Lindblad}, I find that
$\hat{\Cal L}\X{Jp}[\hat \rho(t)] = \Sum_{n \in \mathbb A\X{el}}
\hat{\mathfrak P}^{(n)} \, \hat{\Cal L}\X{Jp}[\hat \rho(t)] \,
\hat{\mathfrak P}^{(n)}$.

The master equation~\eref{eq:mastereqMCWP} in the $n$-electron rigged
Hilbert space~\cite{Madrid:RH-05,Bohm:BS-09} is obtained by projecting
it onto~$\hat{\mathfrak P}^{(n)}$ for~$n \in \mathbb A\X{el}$ writing
\begin{eqnarray}
  \label{eq:projmaster}
  \dfrac{\partial \hat \rho^{(n)}(t)}{\partial t}
    &=& -\imag \, \bigl( \hat H^{(n)\phantom{\dagger}}\X{eff} \,
    \hat \rho^{(n)}(t) - \hat \rho^{(n)}(t) \, \hat H^{(n)\dagger}\X{eff}
    \bigr) \\
  &&{} + \hat{\mathfrak P}^{(n)} \, \hat{\Cal L}\X{Jp}[\hat \rho(t)]
    \, \hat{\mathfrak P}^{(n)} \; , \nonumber
\end{eqnarray}
with~$\hat \rho^{(n)}(t) = \hat{\mathfrak P}^{(n)} \, \hat \rho(t) \,
\hat{\mathfrak P}^{(n)}$.
This implies that
\begin{equation}
  \label{eq:projdens}
  \hat \rho(t) = \Sum_{n \in \mathbb A\X{el}} \hat \rho^{(n)}(t) \; ,
\end{equation}
holds for all times~$t \geq \tau$ where at the
initial time~$\tau$ the system is neutral, \ie, $\hat \rho(\tau)
= \hat \rho^{(N)}(\tau)$.
Let~$\hat{\Cal L}\X{Jp}[\hat \rho(t)]$ exclusively describe quantum jumps
to the same charge state or to higher charge states, \ie,
\begin{equation}
  \label{eq:uporout}
  \hat{\mathfrak P}^{(n)} \, \hat{\Cal L}\X{Jp}[\hat \rho(t)] \,
    \hat{\mathfrak P}^{(n)} = \Sum_{m = n}^N
    \hat{\mathfrak P}^{(n)} \, \hat{\Cal L}\X{Jp}[\hat \rho^{(m)}(t)] \,
    \hat{\mathfrak P}^{(n)} \; .
\end{equation}
With this, Eq.~\eref{eq:projmaster} for~$n = N$ can be solved independently
of higher charge states~$n < N$ because $\hat{\mathfrak P}^{(N)} \,
\hat{\Cal L}\X{Jp}[\hat \rho(t)] \, \hat{\mathfrak P}^{(N)}
= \hat{\mathfrak P}^{(N)} \, \hat{\Cal L}\X{Jp}[\hat \rho^{(N)}(t)]
\, \hat{\mathfrak P}^{(N)}$ provided that $\hat \rho^{(N)}(\tau)$~is
given at the initial time~$\tau$.
For $n < N$, there are terms which couple among rigged Hilbert spaces with
the same or higher electron number.
The number of charge states involved depends on the processes considered,
apart from radiative decay and the coherent interaction in a charge state.
For photoionization and Auger decay, only the charge state with
the next higher electron number is involved.
In the cases that two-electron emission, \eg, due to photoionization shake
off or double Auger decay~\cite{Buth:NX-18,Obaid:FL-18}, are taken
into account, also the charge state with the second next higher electron number
enters Eq.~\eref{eq:projmaster} for~$n \leq N - 2$.
The hierarchy of projected master equations~\eref{eq:projmaster} is
solved successively with numerical methods for differential equations.

\subsection{Random sampling}
\label{sec:MCWP}

The numerical integration of the master equation [\sref{sec:directsum}]
may become prohibitively expensive.
Therefore, I formulate a quantum Monte Carlo~(QMC) algorithm for the solution
of the master equation~\eref{eq:mastereqMCWP} which also is referred
to as random sampling or Monte Carlo wave packets~(MCWPs)~\cite{Dalibard:DP-92,%
Molmer:MC-93,Molmer:MC-96,Plenio:QJ-98,Meystre:QO-99}.
Namely, the system of differential equations for the density
operator~\eref{eq:mastereqMCWP} has number of basis states
squared, \ie, $|\mathbb B|^2$, many equations, if no further
assumptions are made.
Based on the favorable form of Eq.~\eref{eq:mastereqProb}, quantum
trajectories~\cite{Dalibard:DP-92,Molmer:MC-93,Molmer:MC-96,%
Plenio:QJ-98,Meystre:QO-99} can be constructed.
This reduces the task of solving the master equation~\eref{eq:mastereqMCWP}
to the time propagation and quantum-jumping of state vectors.

\subsubsection{Quantum trajectories}
\label{sec:quanttraj}

To construct $|\mathbb T|$~many quantum trajectories, with indices in the
set~$\mathbb T \subseteq \mathbb N$, I start from the
mixture of state vectors~$\mathbb S$ at the initial time~$\tau$ with a
diagonal probabilistic operator~\eref{eq:gendensexp}.
These initial states are the same in all trajectories~$j \in \mathbb T$.
The state vector~$\ket{\Psi_j, t}$ for~$\Psi_j \in \mathbb S$ at
time~$t \geq \tau$ evolves into a state vector at time~$t + \delta t$,
\ie, the next element in all~$\mathbb T$~trajectories, by the following
two rules:

\textbf{First}, \emph{non-Hermitian propagation}~\eref{eq:schromcDeltat}:
with probability~$1 - \delta p_{\Psi_j}(t)$ [Eq.~\eref{eq:normNHprop}], the
state vector at time~$t$ is succeeded at time~$t + \delta t$ by
\begin{equation}
  \label{eq:propagate}
  \ket{\Psi_j, t} \to \dfrac{\ket{\tilde \Psi_j, t + \delta t}}
    {\sqrt{1 - \delta p_{\Psi_j}(t)}} \; .
\end{equation}

\textbf{Second}, \emph{quantum jumps} [last term in Eq.~\eref{eq:mastereqProb}]:
with probability~$\delta p_{\Psi_j}(t)$ [Eq.~\eref{eq:normNHjump}]
the state vector~$\ket{\Psi_j, t}$ at time~$t$ is transformed
into a different state vector at time~$t + \delta t$.
For the transition~$\ket{\psi} \to \ket{\phi}$, this occurs via
\begin{equation}
  \label{eq:jump}
  \ket{\Psi_j, t} \to \sqrt{\dfrac{\delta p_{\Psi_j}(t)}
    {\delta p^{\phi\psi}_{\Psi_j}(t)}}
    \sqrt{\dfrac{\delta t}{\delta p_{\Psi_j}(t)}}
    \hat J_{\phi\psi}^{\phantom{\dagger}} \ket{\Psi_j, t} \; ,
\end{equation}
with the normalization~$\sqrt{\tfrac{\delta p_{\Psi_j}(t)}
{\delta p^{\phi\psi}_{\Psi_j}(t)}}$ and $\delta p^{\phi\psi}_{\Psi_j}(t)
\neq 0$;
otherwise no such jump occurs.
The transition probability from where to where quantum jumps occur
follows from the last line of Eq.~\eref{eq:lossnorm}, and Eqs.~\eref{eq:jumper}
and \eref{eq:normNHjump} to~$\delta p_{\Psi_j}(t) =
\SumInt\limits_{\phi, \psi \in \mathbb B} \delta p^{\phi\psi}_{\Psi_j}(t)$.
This gives the prescription
\begin{equation}
  \label{eq:cascprob}
  \SumInt_{\phi, \psi \in \mathbb B} \dfrac{\delta p^{\phi\psi}_{\Psi_j}(t)}
    {\delta p_{\Psi_j}(t)} = 1 \; ,
\end{equation}
for the conditional probability/probability density~\cite{Arfken:MM-05}
of a quantum jump~$\ket{\psi} \to \ket{\phi}$ to occur
to~$\tfrac{\delta p^{\phi\psi}_{\Psi_j}(t)}{\delta p_{\Psi_j}(t)}$,
if it is known that a quantum jump takes place.

To evaluate the master equation~\eref{eq:mastereqProb}, I
average over quantum trajectories
in conjunction with summing over the probabilistic mixture which
yields the density operator via
\begin{equation}
  \label{eq:trajaverage}
  \hat \rho(t + \delta t) = \Lim_{|\mathbb T| \to \infty}
    \dfrac{1}{|\mathbb T|} \Sum_{j \in \mathbb T}
    \Sum_{\Psi_j \in \mathbb S} \ket{\Psi_j, t + \delta t}
    W_{\Psi_j} \bra{\Psi_j, t + \delta t} \; .
\end{equation}
As Eq.~\eref{eq:trajaverage} is diagonal for~$t + \delta t$ it thus
is for all times~$t \geq \tau$.

\subsubsection{Quantum Monte Carlo algorithm}
\label{sec:QMC}

Let~$\mathbb B$, $\mathbb S$, and $\mathbb T$ be finite.
The QMC algorithm~\cite{Uspenskii:AL-11} to calculate the density
operator~\eref{eq:trajaverage} with~$\mathbb T$~quantum trajectories
proceeds along the following four steps%
\footnote{%
This constitutes a modified algorithm~\cite{Uspenskii:AL-11} over what
is employed in~\cite{Leth:MI-09,Leth:MC-09,Leth:MC-10,Leth:DD-10,%
Leth:DM-11,Leth:DI-11,Jing:MC-16,Jing:LI-16} in which the wave packet
is always propagated whether a quantum jump occurs or not.
Namely, it is decided in the second step whether to propagate or to quantum
jump.
This rearrangement over Refs.~\cite{Leth:MI-09,Leth:MC-09,Leth:MC-10,%
Leth:DD-10,Leth:DM-11,Leth:DI-11,Jing:MC-16,Jing:LI-16} is possible
because the loss of norm of the non-Hermitian propagation~\eref{eq:normNHprop}
can be determined with the state vector at time~$t$ via Eq.~\eref{eq:lossnorm}
and thus the state vector at time~$t + \delta t$ is not required.}
for each trajectory~$j \in \mathbb T$~.

\textbf{First}, \emph{choose} the state vector~$\Psi_j \in \mathbb S$ by
drawing a uniformly distributed random number~$r \in [0; 1[$;
if $r$~falls into the partition
\begin{equation}
  \Sum_{k=1}^{m-1} W_{\Psi^{(k)}} \leq r < \Sum_{k=1}^m W_{\Psi^{(k)}} \; ,
\end{equation}
for an~$m \in \{1, \ldots, |\mathbb S|\}$ where the states in~$\mathbb S$
form the sequence~$\Psi^{(k)} \in \mathbb S$ for~$k \in \{1, \ldots,
|\mathbb S|\}$, and an empty sum is zero, then the state
vector~$\Psi_j = \Psi^{(m)}$ is the initial state of quantum
trajectory~$j$ at time~$\tau$.

\textbf{Second}, \emph{decide} with the probability to quantum
jump~$\delta p_{\Psi_j}(t)$ [Eqs.~\eref{eq:lossnorm}, \eref{eq:normNHjump}]
and a uniformly distributed random number~$r' \in [0; 1]$, whether to go to the
third step for~$r' \in [0; 1 - \delta p_{\Psi_j}(t)[$, or, otherwise,
to the fourth step.

\textbf{Third}, \emph{temporally propagate}~\eref{eq:propagate} the wave packet
using Eq.~\eref{eq:schromcDeltat} without terms~$O(\delta t^2)$.
If the end time of the propagation has not been reached yet,
go to the second step replacing~$t$ by~$t + \delta t$;
otherwise stop.

\textbf{Fourth}, \emph{quantum jump}~\eref{eq:jump} the wave packet.
There are $N\X{I} \in \mathbb N_0$ basis states~$\psi \in \mathbb B$ for
which $\bracket{\psi}{\Psi_j, t} \neq 0$ holds;
they are enumerated as~$\psi^{(\ell)}$ with~$\ell \in \{1, \ldots, N\X{I}\}$.
From these originating states, there are~$N\X{F} \in \mathbb N_0$ basis
states~$\phi \in \mathbb B$ to which a quantum jump goes, \ie, $\exists \ell
\  \gamma_{\phi \psi^{(\ell)}}(t) \neq 0$ [Eq.~\eref{eq:jumper}];
I enumerate them as~$\phi^{(k)}$ with~$k \in \{1, \ldots, N\X{F}\}$.
The pairs of indices of possible quantum jumps are arranged in the
set
\begin{eqnarray}
  \mathbb P &=& \{ (k,\ell) \mid k, \ell \in \mathbb N \land
    1 \leq k \leq N\X{F} \land 1 \leq \ell \leq N\X{I} \nonumber \\
  &&{} \hspace{3.2em} \land \gamma_{\phi^{(k)} \psi^{(\ell)}}(t) \neq 0\} \; .
\end{eqnarray}
The set~$\mathbb P$ is totally ordered~\cite{Skornyakov:OS-16}
under~``$\leq$'' defined for~$(k,\ell), (k',\ell') \in \mathbb P$
by~$(k,\ell) \leq (k',\ell') :\Longleftrightarrow k \, N\X{F} + \ell \leq
k' \, N\X{F} + \ell'$.
The quantum jump~$(k,\ell) \in \mathbb P$ takes place, if the uniformly
distributed random number~$r'' \in [0; 1[$ lies in the interval
\begin{equation}
  \Sum_{\scriptstyle (k',\ell') \in \mathbb P \atop \scriptstyle
    (k',\ell') < (k,\ell)} \!\!\!\!\! \dfrac{\delta p_{\Psi_j}^{\phi^{(k')}
    \psi^{(\ell')}}(t)}{\delta p_{\Psi_j}(t)} \leq r'' < \!\!\!\!\!\!\!\!
    \Sum_{\scriptstyle (k',\ell') \in \mathbb P \atop \scriptstyle
    (k',\ell') \leq (k,\ell)} \!\!\!\!\! \dfrac{\delta
    p_{\Psi_j}^{\phi^{(k')}\psi^{(\ell')}}(t)}{\delta p_{\Psi_j}(t)} \; .
\end{equation}
If the end time of the propagation has not been reached yet,
go to the second step replacing~$t$ by~$t + \delta t$;
otherwise stop.

Due to the first step above, the sum over~$\Psi_j \in \mathbb S$ in
Eq.~\eref{eq:trajaverage} is not required and
$\hat \rho(t) \approx \tfrac{1}{|\mathbb T|}
\Sum_{j \in \mathbb T} \ket{\Psi_j, t} \bra{\Psi_j, t}$ for all~$t \geq \tau$.

\section{Dissociating diatomic molecules}
\label{sec:theory}

I devise a theory for the nuclear dynamics of a diatomic molecule in ultrafast
and intense light that induces (multiple) ionization.
For this purpose, I concretize the abstract formalism
of~\sref{sec:ctqmc}.
Splitting of energy levels due to molecular rotation is assumed
unresolved and I do not consider in detail rotational motion.
In doing so, I acknowledge the fact that such motion takes place on
a much longer, picosecond, time scale compared
with electronic transitions and molecular distortion regarded
here~\cite{Kroto:MR-75,Atkins:MQM-04}.
Specifically, I do not account for excitation of molecular rotations
due to electronic transitions or nuclear distortion and I disregard any
influence on the rotational states by the light.

\subsection{Rigged Hilbert space and basis states}
\label{sec:molhilbert}

The solution of the stationary many-electron Schr\"odinger equation of a
molecule---with all degrees of freedom of the electrons and the nuclei
included---to determine the ground state and excited states is a formidable
task~\cite{Szabo:MQC-89,Monkhorst:CP-87}.
Fortunately, in many cases, the coupling of the motion of the electrons and
the nuclei in a molecule may be neglected because electrons are much
lighter than nuclei and thus the movement of the electrons in
thermal equilibrium is much faster than that of the nuclei
such that the nuclei can be treated as fixed, if only the electronic
structure is to be found.
This leads to a separation of the total Hamiltonian into an electronic
Hamiltonian which depends only parametrically on the
nuclear coordinates and a nuclear Hamiltonian
which contains an electronic potential that averages over the
electronic coordinates and is called Born-Oppenheimer
approximation~(BOA)~\cite{Born:ZQ-27,Szabo:MQC-89,Tannor:IQ-07};
it is assumed throughout.
In BOA, the eigenstates of the stationary Schr\"odinger equation with
the electronic Hamiltonian form potential energy surfaces~(PES)
with respect to the nuclear coordinates~\cite{Szabo:MQC-89,Tannor:IQ-07};
they include also the Coulomb repulsion energy between the nuclei
in their arrangement;
the adiabatic PES are used~\cite{Tannor:IQ-07}.
Thereby, I assume that there is a countable (finite in practice) number of
electronic states which are square integrable, \ie, states in a
Hilbert space (without rigging) which can be achieved by box
normalization~\cite{Merzbacher:QM-98} or, typically, by expanding
the molecular electronic wave functions in terms
of a square-integrable Gaussian basis set~\cite{Szabo:MQC-89}.
Yet there are effects beyond the
BOA~\cite{Koppel:MM-84,Monkhorst:CP-87,Tannor:IQ-07},
not treated here, which shall become relevant for highly excited molecules.

The quantum dynamics of the molecule takes place
in the direct sum rigged Hilbert space introduced abstractly
in~\sref{sec:directsum}.
The space is spanned, as the first component, by the electronic
basis states~$\ket{p ; \, m \, R}{}\X{E}$ from the $p$-electron
Hilbert space for~$p \in \mathbb A\X{el}$ where the subscript~``E''
means that the ket involves the electronic coordinates.
The electronic states are enumerated by~$m \in \{1, \ldots, N\X{st}^{(p)}\}
= \mathbb A\X{st}^{(p)}$ where the number of states considered
is~$N\X{st}^{(p)} \in \mathbb N$.
The dependence of the basis states on the internuclear
distance~$R \in ]0; \infty[$ is only parametrically~\cite{Szabo:MQC-89}.
As a second component, I have the eigenstates of the position operator
in terms of the internuclear distance~$\ket{R}\!{}\X{N}$ where the
subscript~``N'' indicates that the ket is formed with respect to the
nuclear coordinate space.
In total, I have the electronic-distortional basis states given by the tensor
product~\cite{Cohen-Tannoudji:QM-05} of both components for which
orthogonality
\begin{equation}
  \label{eq:elecdist}
  \nuclbracket{p ; \, m \, R}{R}{q ; \, n \, R'}\!{}\X{E} \otimes
    \ket{R'}\!{}\X{N} = \delta(R - R') \, \delta_{pq} \, \delta_{mn}
\end{equation}
holds.
Here $\delta(R - R')$ stands for the Dirac-$\delta$ distribution
and $\delta_{pq}$~is the Kronecker symbol~\cite{Arfken:MM-05}.
There is the completeness relation
\begin{equation}
  \label{eq:completeness}
  \Sum_{p \in \mathbb A\X{el}} \Sum_{\ell \in \mathbb A\X{st}^{(p)}}
    \Int_0^{\infty} \ket{p ; \, \ell \, R} \!{}\X{E} \otimes
    \ket{R}\!{}\X{N} \, {}\X{E}\!\bra{p ; \, \ell \, R} \otimes
    {}\X{N}\!\bra{R} \differential R = \unitop \; .
\end{equation}
A detail concerns the electron-bare molecule which obviously does not have
electronic states.
The internuclear repulsion is described by Coulomb's law.
The corresponding state is denoted as~$\ket{0 ; \, 1 \, R}\!{}\X{E}$;
it is the vacuum (no electron coordinates);
the space are the complex numbers~$\mathbb C$~\cite{Minlos:FS-11}.
The parametric dependence on~$R$ is only included for consistency with
the notation for the other states;
there is none in reality.

\subsection{Nuclear Hamilton operator}
\label{sec:hamiltonian}

The nuclear Hamiltonian that governs the quantum dynamics of the diatomic
molecule in light is
\begin{equation}
  \label{eq:Ham}
  \hat H = \Int_0^{\infty} \bigr( \hat T\X{di} + \hat V\X{el}
    + \hat V\X{li} \bigl) \otimes \ket{R}\!{}\X{N} \, {}\X{N}\!\bra{R}
    \differential R \; ;
\end{equation}
it consists of the nuclear kinetic energy due to distortional
motion~$\hat T\X{di}$, the electronic energy~$\hat V\X{el}$,
and the coherent interaction of the molecule with light~$\hat V\X{li}$.

For a diatomic molecule, the nuclear kinetic energy is
simplified by going into the center of mass reference frame.
Then the two-atom problem is turned into an effective one-atom
problem~\cite{Atkins:MQM-04} with the reduced mass~$\mu$ giving
\begin{equation}
  \label{eq:TnuclOrig}
  -\dfrac{1}{2 \, \mu} \vec \nabla^2
    = -\dfrac{1}{2 \, \mu \, R} \dfrac{\partial^2}{\partial R^2} R
    + B(R) \, \hat{\vec J}^{\,2} \; ,
\end{equation}
in spherical polar coordinates~\cite{Arfken:MM-05} with the internuclear
distance~$R$.
The angular momentum operator~$\hat{\vec J}$ in Eq.~\eref{eq:TnuclOrig}
describes molecular rotation with the principal rotational
constant~\cite{Kroto:MR-75,Atkins:MQM-04}
given by~$B(R) = \tfrac{1}{2 \, \mu \, R^2}$.
The $R$~dependence of~$B(R)$ in Eq.~\eref{eq:TnuclOrig} causes a coupling
of rotational and distortional motion~\cite{Kroto:MR-75}.
I neglect rotational degrees of freedom and write for the nuclear kinetic
energy due to distortional motion
\begin{equation}
  \label{eq:Tnucl}
  \hat T\X{di} = \Sum_{p \in \mathbb A\X{el}} \Sum_{m \in \mathbb
    A\X{st}^{(p)}} \ket{p ; \, m \, R}\!\X{E} \; \Bigl(
    -\dfrac{1}{2 \, \mu \, R} \; \dfrac{\partial^2}{\partial R^2} R \Bigr)' \;
    {}\X{E}\!\bra{p ; \, m \, R} \; .
\end{equation}
The prime on the differential operator indicates that
derivatives of the electronic states with respect to~$R$ are omitted,
\ie, the BOA;
the neglected terms are also local in the internuclear
distance~\cite{Tannor:IQ-07}.

The electronic energy~$\hat V\X{el}$ in Eq.~\eref{eq:Ham} contains all
PESs and can be expressed succinctly by
\begin{equation}
  \label{eq:Vel}
  \hat V\X{el} = \Sum_{p \in \mathbb A\X{el}} \Sum_{m \in
    \mathbb A\X{st}^{(p)}} \ket{p ; \, m \, R}\!\X{E} \; E_m^{(p)}(R) \;
    {}\X{E}\!\bra{p ; \, m \, R} \; ,
\end{equation}
where the PESs are given by~$E_m^{(p)}(R)$.

The coherent interaction of the molecule with light~$\hat V\X{li}$ in
Eq.~\eref{eq:Ham} is treated semiclassically by
\begin{equation}
  \label{eq:Vli}
  \hat V\X{li} = \Sum_{p \in \mathbb A\X{el}} \Sum_{m, n \in \mathbb
    A\X{st}^{(p)}} \ket{p ; \, m \, R}\!\X{E} \;
    \wp^{\;(p)}_{mn}(R, t) \; {}\X{E}\!\bra{p ; \, n \, R} \; .
\end{equation}
Here $\wp^{\;(p)}_{mn}(R, t)$~is the interaction matrix element with the
light that depends on its vector potential~$\vec A(t)$~\cite{Merzbacher:QM-98}
which is given in the laboratory-fixed reference frame.
The polar~$\vartheta$ and azimuth~$\varphi$ angles~\cite{Arfken:MM-05}
specify the orientation of the internuclear axis in the laboratory-fixed
reference frame.
With the Euler matrix~$\mat R\X{E}(\varphi, \vartheta,
0)$~\cite{Kroto:MR-75} I transform~$\vec A(t)$ from the laboratory-fixed frame
to the molecule-fixed frame~$\mat R\X{E}(\varphi, \vartheta, 0) \, \vec A(t)$
(see also Sec.~II~B~1 of Ref.~\cite{Buth:LA-08}).

\subsection{Master equation in Lindblad form}
\label{sec:molmaster}

The abstract equations of~\sref{sec:ctqmc} are rewritten for
diatomic molecules by making the replacements
\begin{eqnarray}
  \label{eq:replacements}
  && \ket{\psi} \longrightarrow \ket{p ; \, m \, R}\!{}\X{E} \otimes
    \ket{R}\!{}\X{N} \nonumber \\
  && \bra{\psi} \longrightarrow {}\X{E}\!\bra{p ; \, m \, R} \otimes
    {}\X{N}\!\bra{R} \nonumber \\
  && \SumInt_{\psi \in \mathbb B} \longrightarrow \Sum_{p \in \mathbb A\X{el}}
    \Sum_{m \in \mathbb A\X{st}^{(p)}} \Int_0^{\infty} \differential R \\
  && \hat{\Cal P}_{\psi} \longrightarrow \hat{\Cal P}_m^{(p)}(R) \nonumber \\
  && \gamma_{\phi \psi}(t) \longrightarrow \delta(R - R') \,
    \gamma_{nm}^{(qp)}(R, t) \nonumber \\
  && \Gamma_{\psi}(t) \longrightarrow \Gamma_m^{(p)}(R, t) \; . \nonumber
\end{eqnarray}
The probabilistic operator in terms of the electronic-distortional
states~\eref{eq:elecdist} reads
\begin{eqnarray}
  \label{eq:densmat}
  \hat \rho(t) &=& \Sum_{p \in \mathbb A\X{el}} \Sum_{m, n \in
    \mathbb A\X{st}^{(p)}} \Int_0^{\infty} \Int_0^{\infty}
    \ket{p ; \, m \, R}\!{}\X{E} \otimes \ket{R}\!{}\X{N} \\
  &&{} \times \rho_{mn}^{(p)}(R, R', t) \, {}\X{E}\!\bra{p ; \, n \, R'}
    \otimes {}\X{N}\!\bra{R'} \differential R \differential R' \nonumber \; ,
\end{eqnarray}
with the density matrix~$\rho_{mn}^{(p)}(R, R', t)$ that
accounts for the form~\eref{eq:projdens}.
The master equation~\eref{eq:mastereqabs} contains incoherent
transitions by photoionization and spontaneous
radiative and electronic decay in~$\hat{\Cal L}\X{Ld}[\hat \rho(t)]$
[Eq.~\eref{eq:Lindblad}].
Hence decoherence results because the emitted photons and electrons
are unobserved which was the initial motivation to use a density operator.
Apart from spontaneous radiative decay, transitions between states with
the same molecular charge are \emph{not} described
by~$\hat{\Cal L}\X{Ld}[\hat \rho(t)]$ [Eq.~\eref{eq:Lindblad}].
Instead, such transitions between PES are treated fully coherently
by~$\hat H$ [Eq.~\eref{eq:Ham}] and thus the quantum mechanical phases
are included.

The Lindblad operator~\eref{eq:Lindblad} is expressed for diatomic
molecules with the projection operator~$\hat{\Cal P}_n^{(q)}(R)$
and the quantum jump operator [Eq.~\eref{eq:jumper}] reading
\begin{equation}
  \label{eq:moljump}
  \hat J_{nm}^{(qp)}(R) = \sqrt{\gamma_{nm}^{(qp)}(R, t)} \> \ket{q ; \, n \, R}
    \!{}\X{E} \otimes \ket{R}\!{}\X{N} \, {}\X{E}\!\bra{p ; \, m \, R}
    \otimes {}\X{N}\!\bra{R} \; .
\end{equation}
Note that a $\delta(R - R')$~factor arising from the
replacements~\eref{eq:replacements} applied to~\eref{eq:jumper}
is not included in the defintion~\eref{eq:moljump}.
The specific form of~\eref{eq:moljump} implies vertical transitions
for which the internuclear distance of the initial and the final state
is the same.
Consequently, the non-Hermitian Hamiltonian~\eref{eq:Vraabs} for the
transitions away from a state~$\ket{p ; \, m \, R}\!{}\X{E} \otimes
\ket{R}\!{}\X{N}$ that cause a loss of population of this state with
the rate~$\Gamma_n^{(p)}(R, t)$ becomes
\begin{equation}
  \label{eq:Vra}
  \hat V\X{ra} = - \dfrac{\imag}{2} \Sum_{p \in \mathbb A\X{el}}
    \Sum_{m \in \mathbb A\X{st}^{(p)}} \Int_0^{\infty} \Gamma_m^{(p)}(R, t) \>
    \hat{\Cal P}_m^{(p)}(R) \differential R \; ,
\end{equation}
where the total transition rate~\eref{eq:MCtotdecra} is
\begin{equation}
  \label{eq:sumGamma}
  \Gamma_m^{(p)}(R, t) = \Sum_{q \in \mathbb A\X{el}}
    \Sum_{n \in \mathbb A\X{st}^{(q)}} \gamma_{nm}^{(qp)}(R, t) \; .
\end{equation}
The quantum jump superoperator~\eref{eq:Lindblad} reads
\begin{widetext}
\begin{eqnarray}
  \label{eq:quantjmp}
  \hat{\Cal L}\X{Jp}[\hat \rho(t)] &=& \Sum_{p, q \in \mathbb A\X{el}}
    \Sum_{m \in \mathbb A\X{st}^{(p)}} \Sum_{n \in \mathbb A\X{st}^{(q)}}
    \Int_0^{\infty} \hat J_{nm}^{(qp)}(R) \; \hat \rho(t) \,
    \hat J_{nm}^{(qp) \, \dagger}(R) \differential R \\
  &=& \Sum_{p, q \in \mathbb A\X{el}} \Sum_{m \in \mathbb A\X{st}^{(p)}}
    \Sum_{n \in \mathbb A\X{st}^{(q)}} \Int_0^{\infty} \gamma_{nm}^{(qp)}(R, t) \>
    \rho_{mm}^{(p)}(R, R, t) \> \hat{\Cal P}_n^{(q)}(R) \differential R
    \; . \nonumber
\end{eqnarray}
\end{widetext}

\subsection{Factored density matrix}
\label{sec:factordens}

In order to derive a matrix representation of the master
equation~\eref{eq:mastereqMCWP} that is amenable to an efficient
solution with methods for PDEs provided that the probabilistic
operator~$\hat \rho(\tau)$ is given at the initial time~$\tau$,
I split~$\hat H^{\vphantom{\prime}}\X{eff}
= \hat T'\X{di} + \hat W$ [Eq.~\eref{eq:HamMC}]
with~$\hat T'\X{di} = \Int_0^{\infty} \hat T\X{di} \otimes
\ket{R}\!{}\X{N} {}\X{N}\!\bra{R} \differential R$ [Eq.~\eref{eq:Tnucl}].
Then the matrix representation of Eqs.~\eref{eq:projmaster},
\eref{eq:projdens}, and \eref{eq:uporout} becomes
\begin{widetext}
\begin{eqnarray}
  \label{eq:fullmatrep}
  \dfrac{\partial \rho_{mn}^{(p)}(R, R', t)}{\partial t} &=&
    \dfrac{\imag}{2 \mu} \Bigl( \dfrac{1}{R} \dfrac{\partial^2}{\partial R^2} R
    - \dfrac{1}{R'} \dfrac{\partial^2}{\partial R'^2} R' \Bigr) \,
    \rho_{mn}^{(p)}(R, R', t)
    - \imag \Sum_{\ell \in \mathbb A\X{st}^{(p)}} W_{m\ell}^{(p)}(R, t)
    \, \rho_{\ell n}^{(p)}(R, R', t) \\
  &&{} + \imag \Sum_{\ell \in \mathbb A\X{st}^{(p)}} \rho_{m\ell}^{(p)}
    (R, R', t) \, W_{n\ell}^{(p)\,*}(R', t)
    + \delta(R - R') \, \delta_{mn} \Sum_{q = p}^N \Sum_{\ell \in \mathbb
    A\X{st}^{(q)}} \gamma_{m\ell}^{(pq)}(R, t) \,
    \rho_{\ell\ell}^{(q)}(R, R, t) \; , \nonumber
\end{eqnarray}
\end{widetext}
where the first summand on the right-hand side stems from the
Hermitian~$\hat T'\X{di}$.
Moreover, I inserted the completeness relation~\eref{eq:completeness}
to transform the operator products~$\hat W \,
\hat \rho(t)$ and $\hat \rho(t) \, \hat W^{\dagger}$
in~\eref{eq:mastereqMCWP} giving the second and third summands
in~\eref{eq:fullmatrep};
the matrix elements of~$\hat W$ are
\begin{eqnarray}
  \label{eq:Wmatel}
  && {}\X{E}\!\bra{p ; \, m \, R} \otimes {}\X{N}\!\bra{R} \hat W
    \ket{q ; \, n \, R'}\!{}\X{E} \otimes \ket{R'}\!{}\X{N} \nonumber \\
  &=& \delta(R - R') \, \delta_{pq} \, W_{mn}^{(p)}(R, t) \\
  &=& \delta(R - R') \, \delta_{pq} \, \Bigl[ \delta_{mn} \, \Bigl(
    E_m^{(p)}(R) - \dfrac{\imag}{2} \, \Gamma_m^{(p)}(R, t) \Bigr) \nonumber \\
  &&{} + \wp^{\;(p)}_{mn}(R, t) \Bigr] \; . \nonumber
\end{eqnarray}
The fourth summand in~\eref{eq:fullmatrep} stems from the matrix
elements~${}\X{E}\!\bra{p ; \, m \, R} \otimes {}\X{N}\!\bra{R}
\hat{\Cal L}\X{Jp}[\hat \rho(t)] \ket{q ; \, n \, R'} \!{}\X{E}
\otimes \ket{R'}\!{}\X{N}$ of the quantum
jump superoperator~\eref{eq:quantjmp} considering~\eref{eq:uporout}
to restrict the sum over charge states.

\Eref{eq:fullmatrep} could, in principle be used---by discretizing
the continuous variables~$R$ and $R'$---to
calculate~$\rho_{mn}^{(p)}(R, R', t)$.
Yet this leads to an enormously high-dimensional matrix equation
which is in practice intractable for all but the simplest systems.
But $\hat H$ does not couple states with different internuclear distance
and neither do the transition rates in~$\hat{\Cal L}\X{Ld}[\hat \rho(t)]$
[Eqs.~\eref{eq:Lindblad}, \eref{eq:Vra}, and \eref{eq:quantjmp}], \ie,
all matrix representations but~$\rho_{mn}^{(p)}(R, R', t)$ are local
in the internuclear distance.
Hence I make the product ansatz
\begin{equation}
  \label{eq:factordens}
  \rho_{mn}^{(p)}(R, R', t) = \dfrac{\varsigma_{mn}^{(p)}(R, t)}{R} \>
    \dfrac{\varsigma_{mn}^{(p)\,*}(R', t)}{R'} \; ,
\end{equation}
for the density matrix.
Substituting this into expression~\eref{eq:fullmatrep} leads---by multiplying
with~$R \, R'$, integrating over~$R'$, and dividing
by~$\tilde\varsigma_{mn}^{(p)\,*}(t) = \Int_0^{\infty} \varsigma_{mn}^{(p)\,*}
(R', t) \differential R'$, if nonzero---to the decoupled equation
\begin{widetext}
\begin{eqnarray}
  \dfrac{\partial \varsigma_{mn}^{(p)}(R, t)}{\partial t} &=&
    \dfrac{\imag}{2 \mu} \dfrac{\partial^2}{\partial R^2} \,
    \varsigma_{mn}^{(p)}(R, t) + U_{mn}^{(p)}(R, t)
     - \imag \Sum_{\ell \in \mathbb A\X{st}^{(p)}} W_{m\ell}^{(p)}(R, t)
    \, \varsigma_{\ell n}^{(p)}(R, t) \\
  &&{} + \dfrac{\delta_{mn}}{\tilde\varsigma_{mm}^{(p)\,*}(t)}
    \Sum_{q = p}^N \Sum_{\ell \in \mathbb
    A\X{st}^{(q)}} \gamma_{m\ell}^{(pq)}(R, t) \,
    |\varsigma_{\ell\ell}^{(q)}(R, t)|^2 \; , \nonumber
\end{eqnarray}
with the complex potential
\begin{eqnarray}
  U_{mn}^{(p)}(R, t) &=& -\dfrac{1}{\tilde\varsigma_{mn}^{(p)\,*}(t)} \Bigl[
    \varsigma_{mn}^{(p)}(R, t) \Bigl( \dfrac{\partial \tilde\varsigma_{mn}
    ^{(p)\,*}(t)}{\partial t}
    + \dfrac{\imag}{2 \mu} \Int_0^{\infty} \dfrac{\partial^2}{\partial R'^2} \,
    \varsigma_{mn}^{(p)\,*}(R', t) \differential R' \Bigr) \\
  &&{} - \imag \Sum_{\ell \in \mathbb A\X{st}^{(p)}}
    \varsigma_{m\ell}^{(p)}(R, t) \Int_0^{\infty} \varsigma_{m\ell}^{(p)\,*}
    (R', t) \>W_{n\ell}^{(p)\,*}(R', t) \differential R' \Bigr] \; . \nonumber
\end{eqnarray}
\end{widetext}
The factorization of the density matrix~\eref{eq:factordens} has broken down
the problem of calculating~$\varsigma_{mn}^{(p)}(R, t)$ to a equation
that only depends on a single internuclear distance which is numerically
tractable.

\subsection{Quantum Monte Carlo formalism}
\label{sec:molQMC}

Here I specialize the quantum Monte Carlo algorithm from~\sref{sec:MCWP}
to dissociating molecules.
Given a set~$\mathbb S$ of states at the initial time~$\tau$,
non-Hermitian propagation for~$\Psi \in \mathbb S$ is governed by equations
of motion~(EOMs) which are obtained within the framework of MCWP from the
time-dependent Schr\"odinger equation~\eref{eq:schromc} with the effective
Hamiltonian~\eref{eq:HamMC} by making the ansatz
\begin{equation}
  \label{eq:wavepacket}
  \ket{\Psi, t} = \Sum_{q \in \mathbb A\X{el}} \Sum_{n \in \mathbb
    A\X{st}^{(q)}} \Int_0^{\infty} \dfrac{\varphi_n^{(q)}(R, t)}{R}
    \ket{q ; \, n \, R}\!{}\X{E} \otimes \ket{R}\!{}\X{N} \differential R \; ,
\end{equation}
for the wave packet of the electronic-distortional motion.
The expansion coefficients are the nuclear wave
functions~$\tfrac{\varphi_n^{(q)}(R, t)}{R}$ where a division
by~$R$ is made explicitly in order to simplify the resulting
EOMs for the nuclear dynamics.
The EOMs for the quantum dynamics in a specific charge state
follow~\cite{Leth:MI-09,Leth:MC-09,Leth:MC-10,Leth:DD-10,%
Leth:DM-11,Leth:DI-11,Jing:MC-16,Jing:LI-16} by inserting the wave
packet~\eref{eq:wavepacket} and
the effective Hamiltonian~\eref{eq:HamMC}, \eref{eq:Ham},
and \eref{eq:Vra} into the time-dependent Schr\"odinger
equation~\eref{eq:schromc} and projecting onto the basis
state~${}\X{E}\!\bra{p ; \, m \, R} \otimes {}\X{N}\!\bra{R}$ giving
\begin{widetext}
\begin{eqnarray}
  \label{eq:EOMs}
  \imag \, \dfrac{\partial}{\partial t} \varphi_m^{(p)}(R, t)
    &=& \Sum_{q \in \mathbb A\X{el}} \Sum_{n \in \mathbb
    A\X{st}^{(q)}} \Int_0^{\infty} {}\X{E}\!\bra{p ; \, m \, R}
    \otimes {}\X{N}\!\bra{R} R \> \hat H\X{eff} \,
    \dfrac{\varphi_n^{(q)}(R', t)}{R'} \ket{q ; \, n \, R'}\!{}\X{E} \otimes
    \ket{R'}{}\X{N} \differential R' \\
  &=& \Bigl[ -\dfrac{1}{2\,\mu} \dfrac{\partial^2}{\partial R^2}
    + E_m^{(p)}(R) - \dfrac{\imag}{2} \> \Gamma_m^{(p)}(R, t) \Bigr]
    \varphi_m^{(p)}(R, t) - \Sum_{n \in \mathbb A\X{st}^{(q)}}
    \wp^{\;(p)}_{mn}(R, t) \> \varphi_n^{(p)}(R, t) \; . \nonumber
\end{eqnarray}
\end{widetext}
The EOMs~\eref{eq:EOMs} form a linear system of PDEs~\cite{Arfken:MM-05}
for the nuclear wave functions
which are first-order in the time derivative and second-order in the
spatial derivative.
Prior to the occurrence of any quantum jumps in the QMC~algorithm
of~\sref{sec:QMC}, the summand in the trajectory-average for the density
operator [confer Eq.~\eref{eq:trajaverage}] is the outer
product~$\ket{\Psi, t} \bra{\Psi, t}$ of~\eref{eq:wavepacket} describing
only the neutral molecule.
The probability for quantum jumping is then derived from~\eref{eq:lossnorm}
via~\eref{eq:replacements} to
\begin{equation}
  \label{eq:initialjump}
  \delta p_{\Psi, nm}^{(qp)}(t) = \delta t \Int_0^{\infty}
    \gamma_{nm}^{(qp)}(R, t) \, \bigl| {}\X{E}\!\bra{p ; \, m \, R}
   \otimes {}\X{N}\!\left<\right.\!R\left.\right|\Psi, t
    \!\left.\right> \bigr|^2
    \differential R \; .
\end{equation}
The expansion coefficients are given by~$\nuclbracket{p ; \, m \, R}{R}
{\Psi, t} = \tfrac{\varphi_m^{(p)}(R, t)}{R}$.
Note the integration over~$R$ in the equation;
for a direct translation with~\eref{eq:replacements}, it would be
missing.
Doing so is reminiscent of Eq.~\eref{eq:intjump} below and acknowledges
that quantum jumps shall only take place between discrete indices.

Quantum jumping is mediated by the operator~$\hat J_{nm}^{(qp)}(R)$
[Eq.~\eref{eq:moljump}] which enters the formulation of the
superoperator~$\hat{\Cal L}\X{Jp}[\hat \rho(t)]$ [Eq.~\eref{eq:quantjmp}].
Yet this straight translation of the quantum jump operator
via~\eref{eq:replacements} to diatomic molecules is unfavorable
because it depends on the continuous variable~$R \in ]0 ; \infty[$.
In other words, the $\hat J_{nm}^{(qp)}(R)$ promote a Dirac-$\delta$
wave packet.
Upon discretization, this implies that there are a huge number of quantum
jump operators.
To circumvent this situation, I define an integrated quantum jump
operator over the internuclear distance via
\begin{equation}
  \label{eq:intjump}
  \hat{\mathfrak J}_{nm}^{(qp)} = \Int_0^{\infty} \hat J_{nm}^{(qp)}(R)
    \differential R \; .
\end{equation}
This trick removes the dependance of the original quantum jump
operator~$\hat J_{nm}^{(qp)}(R)$ on the continuous variable~$R$ and
thus only a dependance of~$\hat{\mathfrak J}_{nm}^{(qp)}$ on
discrete variables remains.
However, the definition~\eref{eq:intjump} is not motivated by the
structure of the equations derived so far and I need to examine
how this operator can be put into the expressions to remove
their dependence on~$\hat J_{nm}^{(qp)}(R)$.
\Eref{eq:intjump} is introduced in Refs.~\cite{Leth:MI-09,Leth:MC-09,%
Leth:MC-10,Leth:DD-10,Leth:DM-11,Leth:DI-11,Jing:MC-16,Jing:LI-16}
but not motivated and using the form~\eref{eq:intjump} has consequences
unaccounted for therein.
Namely, inspecting the first line in~\eref{eq:quantjmp}, I realize that
the quantum jump superoperator in terms of the~$\hat{\mathfrak J}_{nm}^{(qp)}$
has to be expressed as
\begin{eqnarray}
  \label{eq:intjumpLjp}
  \hat{\Cal L}\X{Jp}[\hat \rho(t)] &=& \Sum_{p, q \in \mathbb A\X{el}}
    \Sum_{m \in \mathbb A\X{st}^{(p)}} \Sum_{n \in \mathbb A\X{st}^{(q)}}
    \Int_0^{\infty} \hat Q(R) \, \hat{\mathfrak J}_{nm}^{(qp)} \\
  &&{} \times \hat \rho(t) \, \hat{\mathfrak J}_{nm}^{(qp) \, \dagger}
    \, \hat Q(R) \differential R \; , \nonumber
\end{eqnarray}
with~$\hat Q(R) = \unitop\X{E} \otimes \ket{R}\!{}\X{N} \, {}\X{N}\!\bra{R}
= \hat Q^{\dagger}(R)$, where $\unitop\X{E}$~is the unit operator in
the direct sum Hilbert space of the electronic states.
Clearly, I have $\hat Q(R) \, \hat{\mathfrak J}_{nm}^{(qp)}
= \hat J_{nm}^{(qp)}(R)$ and Eq.~\eref{eq:intjumpLjp} implies
that~$\hat{\mathfrak J}_{nm}^{(qp)} \, \ket{\Psi, t} \bra{\Psi, t} \,
\hat{\mathfrak J}_{nm}^{(qp) \, \dagger}$ is not a term in the sums therein.
Put differently, expressing~\eref{eq:quantjmp} by
replacing~$\hat J_{nm}^{(qp)}(R)$ with~$\hat{\mathfrak J}_{nm}^{(qp)}$ gives
incorrect equations.

A quantum jump mediated by~\eref{eq:intjump} can no longer be expressed
as an operation on a state vector as before~\eref{eq:jump}
but has to be written as a density operator.
Let the state vector at~$t + \delta t$ for fixed quantum
numbers~$p,q \in \mathbb A\X{el}$, $m \in \mathbb A\X{st}^{(p)}$, and
$n \in \mathbb A\X{st}^{(q)}$, be the promoted wave packet
\begin{eqnarray}
  \label{eq:qujpwave}
  \ket{\Psi, t + \delta t} &=& \sqrt{\tfrac{\delta t}{\delta p_{\Psi,
    nm}^{(qp)}(t)}} \; \hat{\mathfrak J}_{nm}^{(qp)} \, \ket{\Psi, t} \\
  &=& \sqrt{\tfrac{\delta t}{\delta p_{\Psi, nm}^{(qp)}(t)}} \Int_0^{\infty}
    \sqrt{\gamma_{nm}^{(qp)}(R, t)} \; \dfrac{\varphi_m^{(p)}(R, t)}{R} \;
    \nonumber \\
  &&{} \times \ket{q ; \, n \, R}\!\X{E} \otimes \ket{R}\!\X{N} \differential R
    \; , \nonumber
\end{eqnarray}
which is normalized, \ie, $\bracket{\Psi, t + \delta t}
{\Psi, t + \delta t} = 1$.
With this the probabilistic operator reads
\begin{equation}
  \label{eq:promdens}
  \hat \varrho(t + \delta t) = \Int_0^{\infty} \hat Q(R) \,
    \ket{\Psi, t + \delta t} \bra{\Psi, t + \delta t} \hat Q(R)
    \differential R \; ;
\end{equation}
its trace is unity, \ie, $\Tr \hat \varrho(t + \delta t)
= \bracket{\Psi, t + \delta t}{\Psi, t + \delta t} = 1$

The promoted wave packet~\eref{eq:qujpwave} is the basis for the next cycle
in the Monte Carlo algorithm.
The~$\hat \varrho(t + \delta t)$ is
the result of a quantum jump occurring at~$t$.
For the next time step from~$t + \delta t$ to~$t + 2 \, \delta t$,
I insert~$\hat \varrho(t + \delta t)$
into the short-time master equation~\eref{eq:mastereqDeltat}:
\begin{widetext}
\begin{eqnarray}
  \label{eq:masterproj}
  \hat \varrho(t + 2 \, \delta t) &=& \hat \varrho(t + \delta t)
    - \imag \, \delta t \Int_0^{\infty} \hat Q(R) \,
    \bigl( \hat H\X{eff} \, \ket{\Psi, t + \delta t} \bra{\Psi, t + \delta t}
    - \ket{\Psi, t + \delta t} \bra{\Psi, t + \delta t} \,
    \hat H^{\dagger}\X{eff} \bigr) \, \hat Q(R) \differential R \\
  &&{}+ \delta t \Int_0^{\infty} \hat Q(R) \> \Bigl( \Sum_{p, q \in \mathbb
    A\X{el}} \Sum_{m \in \mathbb A\X{st}^{(p)}} \Sum_{n \in \mathbb
    A\X{st}^{(q)}} \hat{\mathfrak J}_{nm}^{(qp)} \ket{\Psi, t + \delta t}
    \bra{\Psi, t + \delta t} \, \hat{\mathfrak J}_{nm}^{(qp) \, \dagger}
    \Bigr) \> \hat Q(R) \differential R + O(\delta t^2) \; . \nonumber
\end{eqnarray}
\end{widetext}
Here I exploit the commutators~$[\hat H^{\phantom{\dagger}}\X{eff}, \hat Q(R)]
= [\hat H^{\dagger}\X{eff}, \hat Q(R)] = 0$.
Further inserting~\eref{eq:promdens} into~\eref{eq:intjumpLjp}
is the same as using the outer product instead;
the superoperator~\eref{eq:intjumpLjp} already contains a projection
with~$\hat Q(R)$.
Thereby, I use~$\hat Q(R) \, \hat Q(R') = \delta(R - R') \, \hat Q(R)$ and
the commutators~$[\hat{\mathfrak J}_{nm}^{(qp)}, \hat Q(R)]
= [\hat{\mathfrak J}_{nm}^{(qp) \dagger}, \hat Q(R)] = 0$
and replace $[\delta(R - R')]^2$ by~$\delta(R - R')$.
\Eref{eq:masterproj} corresponds to a solution of~\eref{eq:mastereqDeltat}
with~$\ket{\Psi, t + \delta t} \bra{\Psi, t + \delta t}$ and a
projection of its result at time~$t + 2 \, \delta t$
with~$\hat Q(R)$ thereafter, \ie,
performing an operation as in~\eref{eq:promdens}.
Based on this finding, I use state vectors as before and either
propagate~\eref{eq:EOMs} or quantum jump~\eref{eq:intjumpLjp};
the result, however, then needs to be projected~\eref{eq:promdens}
to be meaningful.

The probabilities for quantum jumping---after an initial quantum jump
has occurred---need to be derived for the
probabilistic operator~\eref{eq:promdens}.
I use the trace condition~\eref{eq:tracecond} for the quantum jump
superoperator~\eref{eq:intjumpLjp} giving
\begin{equation}
  \delta p_{\Psi}(t) = \delta t \; \Tr \hat{\Cal L}\X{Jp}[\hat
    \varrho(t)] = \Sum_{p, q \in \mathbb A\X{el}} \Sum_{m \in
    \mathbb A\X{st}^{(p)}} \Sum_{n \in \mathbb A\X{st}^{(q)}}
    \delta p_{\Psi,nm}^{(qp)}(t) \; ;
\end{equation}
This leads to the \emph{same} expression~\eref{eq:initialjump}
for~$\delta p_{\Psi,nm}^{(qp)}(t)$ as before.

The upshot is that one can solve the master equation by Monte Carlo
wave packet propagation with~\eref{eq:intjump} along the lines
of~\sref{sec:QMC} where, however, a projection~\eref{eq:promdens} is
required henceforth after the first quantum jump has occurred.
Let me go through the algorithm of~\sref{sec:QMC} in detail.
The first step selects the starting state vector and is unmodified.
The second step uses the probability~\eref{eq:initialjump}
to decide whether to quantum jump or not.
The third step uses~\eref{eq:EOMs} to temporally propagate.
The fourth step quantum jumps employing~\eref{eq:qujpwave}.
The density operator of the system is then approximated analogously
to~\eref{eq:trajaverage} where, however, the outer products
are replaced by~\eref{eq:promdens} once a quantum jump has happened.

\section{Conclusion}
\label{sec:conclusion}

Molecules in ultrafast and intense light exhibit a fascinating phenomenology
of intertwined electronic and nuclear quantum dynamics involving
(multiple) ionization, radiative and electronic decay, rotation,
vibration and dissociation.
To tackle this situation, I devise a rigorous quantum optical formalism,
initially in abstract form, centered around the master equation in
Lindblad form.
It describes the time evolution of the probabilistic operator
for system-reservoir interaction using the Markoff approximation
allowing for the dissipation of particles and energy.
The system, thereby, is the dissociating molecule which dissipates
photons and electrons into the reservoir, \ie, the continuum.
A short-time approximation of the master equation is derived.
In a direct sum of rigged Hilbert spaces for multiple charge states,
I reduce it to a hierarchy of equations.
This allows me to solve them using methods for systems
of differential equations.
Alternatively, quantum Monte Carlo wave packet propagation~(MCWP)
is proposed that yields the density operator by averaging over
an ensemble of quantum trajectories.
The MCWP approach is computationally more efficient than a numerical integration
of the master equation and facilitates a lucid physical interpretation of
the resulting quantum trajectories in terms of propagation and
quantum jumps for a single state vector.
The abstract formalism is concretized to diatomic molecules in
Born-Oppenheimer approximation.
The basis states are the electronic-distortional states for a fixed-in-space
intermolecular axis, \ie, molecular rotation is disregarded.
An equation for a factored density matrix which depends only on a
single internuclear distance is derived.
It is amenable to a solution with methods for PDEs.
The MCWP formulation requires to put special emphasize on quantum jumps
whereby an integrated jump operator is used.
This, however, necessitates the use of a projection afterwards which has
incorrectly been omitted in Refs.~\cite{Leth:MI-09,Leth:MC-09,Leth:MC-10,%
Leth:DD-10,Leth:DM-11,Leth:DI-11,Jing:MC-16,Jing:LI-16}, \ie, a
density operator description is necessay.
Forthcoming computational work based on the present formalism shall
uncover the impact of the correct treatment of quantum jumps put forward
here in comparison to Refs.~\cite{Leth:MI-09,Leth:MC-09,Leth:MC-10,%
Leth:DD-10,Leth:DM-11,Leth:DI-11,Jing:MC-16,Jing:LI-16}.

Deterministic sampling has been proposed to solve the master
equation for a small number of potential energy surfaces~(PES)
considered~\cite{Leth:MI-09,Leth:MC-09,Leth:MC-10,Leth:DD-10,%
Leth:DM-11,Leth:DI-11,Jing:MC-16,Jing:LI-16}.
This is a complementary notion to the solution of the hierarchy of
equations for different charge states put forward here.
In the specialized case of no coupling of electronic states with the same
number of electrons by the quantum jump part of the Lindblad
superoperator---that is exclusively considered in
Refs.~\cite{Leth:MI-09,Leth:MC-09,Leth:MC-10,Leth:DD-10,%
Leth:DM-11,Leth:DI-11,Jing:MC-16,Jing:LI-16}---probability theory can be used.
Starting from a propagation on a single ground-state PES
quantum jumps to other states are made in sufficient frequency
in order to sample these transitions adequately for the
quantities of interest to be converged, \eg, the kinetic energy release.
This prescription is a formally exact probabilistic solution.
However, there is also a weakness of a strong increase
of the number of possible pathways.
Namely, the ground-state wave packets need to be propagated for all times and
quantum jumps are applied to all accessible destinations
with certain frequency from whence a further propagation is necessary.
In the light of this, I conclude that deterministic sampling
does not seem to provide any advantage over the numerical integration
of the master equation.
On the contrary, only a single solution of the PDEs is required whereas
multiple propagations are required for higher charge states in
deterministic sampling.

The presented quantum optical formalism is very basic and may find
ubiquitous use to describe the quantum dynamics of light-matter interaction.
It thus provides manifold prospects for future research.
A particular feature of the presented approach is that ionization
processes which reduce the number of electrons are accounted
for solely by transition rates and the outgoing electron itself
is not regarded ``it just disappears.''
If also the electron dynamics shall be included, some modifications
are in order using a complex absorbing potential to absorb the
electron~\cite{Selsto:AB-10}.

So far the MCWP method has only been applied in the context of the
photoionization of diatomic molecules~\cite{Leth:MI-09,Leth:MC-09,%
Leth:MC-10,Leth:DD-10,Leth:DM-11,Leth:DI-11,Jing:MC-16,Jing:LI-16}
in the optical and the \XUV~regime.
An extension to the \xray~domain, where radiative and electronic
decay become manifest~\cite{Hoener:FA-10,Fang:MI-12,Buth:UA-12,Liu:RE-16}, is a
fascinating and highly pressing issue.
Thereby, not only electronic decay involving core holes may be relevant
but there are also situations in which inner-valence vacancies undergo
an ultrafast electronic decay~\cite{Buth:IM-03,Scheit:IP-03}.

The formalism of~\sref{sec:ctqmc} and \sref{sec:theory} is also directly
applicable to atoms.
To this end, I simply omit the internuclear distance in the basis states
and modify the equations appropriately.
The MCWP is a step ahead over a Monte Carlo solution of
rate equations for atoms~\cite{Son:MC-12,Son:EM-15} as also coherent
phenomena are describeable.

Molecular rotation is not described in this work because it only leads to
a small energy splitting compared with the electronic-distortional motion.
In future work, one may consider a treatment of molecular rotations,
specifically rotational excitation by electronic transitions and nuclear
distortion and molecular alignment by the light~\cite{Buth:LA-08}.
Here the consequences of the nuclear expansion (flexible
rotor)~\cite{Kroto:MR-75} of a dissociating molecule on the rotation
spectra is to be treated thoroughly.

\Xray~quantum optics~\cite{Adams:QO-13} comes into reach with molecules;
a number of studies have been conducted for atoms which
now await an examination for molecules facilitated by the present
article.
This bears a high potential for discovery that goes beyond what
can be achieved with rate equations~\cite{Buth:UA-12,Liu:RE-16,%
Buth:NX-18,Obaid:FL-18,Buth:LT-19}.
Namely, if the \xray~energy is tuned to a resonance, then coherences
manifest.
Especially the two-color physics of FEL x~rays and an optical laser
offers further promising avenues~\cite{Adams:QO-13}.

I have been mostly concerned with \XUV~light and x~rays from FELs here but
attosecond light sources offer exciting new possibilities for studying
molecules and the presented theory shall be applicable in this
situation as well.
Such sources are particularly interesting because they offer a much
finer control of the beam characteristics compared with present-day FELs.

\section*{Acknowledgments}

I am grateful to Mathias Nest for thoughtful discussions.
So I am to Lorenz S.~Cederbaum and Jochen Schirmer and
I thank them for a critical reading of the manuscript.
This research did not receive any specific grant from funding
agencies in the public, commercial, or not-for-profit sectors.
Declarations of interest: none.

\end{document}